\documentclass[aps,showpacs,amsfonts,nofootinbib,preprintnumbers,nobalancelastpage]{revtex4}

\usepackage[english]{babel}
\usepackage{amsmath,amssymb}
\usepackage[dvips]{graphicx}
\usepackage[sort&compress]{natbib}
\usepackage{bbm}
\usepackage{color}
\usepackage{ulem}

\DeclareMathAlphabet{\mathsc}{OT1}{cmr}{m}{sc}

\allowdisplaybreaks

\def\321{SU(3) $\otimes$ SU(2) $\otimes$ U(1)}

\def\lsim{\raise0.3ex\hbox{$\;<$\kern-0.75em\raise-1.1ex\hbox{$\sim\;$}}}
\def\gsim{\raise0.3ex\hbox{$\;>$\kern-0.75em\raise-1.1ex\hbox{$\sim\;$}}}

\newcommand{\flux}[2][]{\ensuremath{\ifthenelse{\equal{#1}{}}{}{^{#1}\!}\mathit{#2}}}

\begin{document}

\preprint{$\;$}

\title{Displaced vertices in GMSB models at LHCb}

\author{F.\ de Campos}
\email{fernando.carvalho@ict.unesp.br}
\affiliation{Department of Environmental Engineering, 
  Institute of Science and Technology, Universidade Estadual Paulista, 
  S\~ao Jos\'e dos Campos -- SP,  Brazil }

\author{M.\ B.\ Magro}
\email{magro@fma.if.usp.br}
\affiliation{Instituto de F\'{\i}sica,
             Universidade de S\~ao Paulo, S\~ao Paulo -- SP, Brazil.}
\affiliation{Centro Universit\'ario Funda\c{c}\~ao Santo Andr\'e,
             Santo Andr\'e -- SP, Brazil.}

\begin{abstract}

  We consider minimal Gauge Mediated Supersymmetry Breaking models
  at which the next to lightest supersymmetric
  particle is a neutralino with a large
  enough decay length to be detected at the CERN Large Hadron Collider.
  We analyze the potential of the LHCb experiment to 
  determine the discovery reach for such models and found that the LHCb will 
  be able to probe such models up to the energy breaking scale of 
  $\Lambda = 130$ TeV.

\end{abstract}


\maketitle

\section{Introduction}

Elucidating the electroweak breaking sector of the Standard Model (SM)
constitutes a major challenge for the Large Hadron Collider (LHC) at
CERN. Supersymmetry provides an elegant way to stabilize the Higgs
boson scalar mass against quantum corrections provided supersymmetric
states are not too heavy, with some of them expected within reach for
the LHC.  Searches for supersymmetric particles constitute an important
item in the LHC agenda~\cite{Chatrchyan:2011zy,daCosta:2011qk,%
ATLAS:2011ad,Aad:2011qa,Aad:2011cwa,Aad:2011zj,Khachatryan:2011tk,%
Chatrchyan:2011bz,Chatrchyan:2011wba,Chatrchyan:2011ff}, as
many expect signs of supersymmetry (SUSY) to be just around the
corner and, indeed, some SUSY searches has been done recently by the ATLAS 
experiment \cite{atlas-conf140}.  However, the first searches up to 
$\sim$ 5 fb$^{-1}$ at the
LHC interpreted within specific frameworks, such as Constrained
Minimal Supersymmetric Standard Model (CMSSM) or minimal supergravity
(mSUGRA) indicate that squark and gluino masses are in excess of $\sim
1$ TeV \cite{Bechtle:2012zk}. \smallskip

Despite intense efforts over more than thirty years, little is known
from first principles about how exactly to realize or break
supersymmetry.  As a result one should keep an open mind as to which
theoretical framework is realized in nature, if any. Here, we consider  
gauge mediated supersymmetry breaking models (GMSB) in which SUSY is broken in 
a hidden sector by messanger fields which interacts with the SM sector through
gauge mediated interactions \cite{gmsb:models,gmsb:review}. In these scenarios, 
the lightest supersymmetrical
particle (LSP)is the gravitino, presenting a mass below the keV scale, while the 
next-to-lightest supersymmetrical particle (NLSP) is, most of the time, the 
lightest neutralino. Since, in most of the parameter space spectrum, the 
coupling between the gravitino and the lightest neutralino is small, the decay 
length of the NLSP is usually large and it may produce displaced vertices that 
can be detected at LHC.

The LHCb experiment consists of a front-end detector mainly designed to 
investigate b-physcis and, therefore, is very sensitive to performing displaced 
vertices searches. In this work, we consider the capability of the LHCb to detect
displaced vertices coming from the lightest neutralino decays in the framework 
of GMSB models. The paper is organized as follows. In Sec.~\ref{sec:model}, we 
present a brief resume of the characteristcs of GMSB models, pinpointing the 
large decay length of the NLSP. In Sec~\ref{ana:frame}, we show the details of our
signal simulation and the analysis of the results. Finally, in 
Sec.~\ref{sec:conclusions}, we draw our conclusions and comments regarding recent 
ATLAS and CMS higgs search results.

\smallskip

\section{Gauge Mediated Supersymmetry Breaking models}
\label{sec:model}

Gauge Mediated Supersymmetry Breaking models are well motivated, since they can 
solve the SM hierarchy problem by the introduction of supersymmetric partners, 
as well as the SUSY flavour problem by the gauge mediation. GMSB models assume 
that SUSY is dinamically broken by the introduction of messanger fields in a 
hidden sector where they couple with the SM fields through gauge mediated 
interactions. In this way, such GMSB models are defined by six parameters
\begin{equation}
\Lambda\,,\, M_m\,,\, \tan\beta\,,\, {\mathrm{sign}}(\mu)\,,\,
N_5\,,\, C_{grav}\,,
\end{equation}
where $\Lambda$ is the energy scale of the breaking, $M_m$ is the
messenger mass scale, $\mu$ is the Higgs mass term, $\tan\beta$ is the ratio 
between the Higgs field vacuum expectation values (vevs), $N_5$ is the
number of messenger chiral supermultiplets and $C_{grav}$ is the scale factor of the 
gravitino mass. In the minimal realization of GMSB
we take  $N_5 = 1$ and $C_{grav} = 1$.\smallskip

\begin{figure}
\includegraphics[width=0.49\textwidth,angle=0]{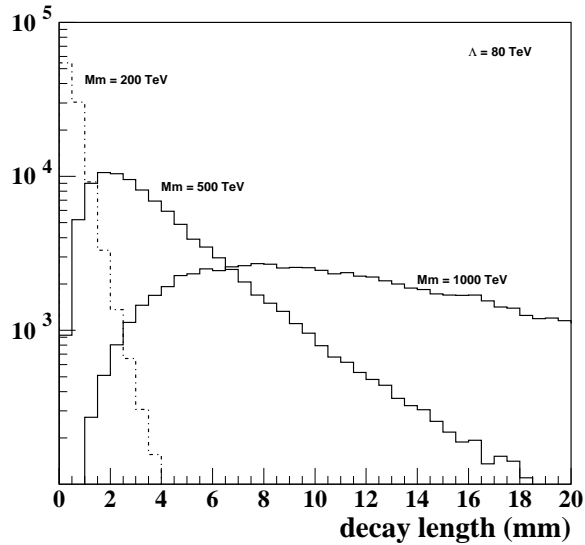}
\caption{Lightest neutralino decay length for several values $M_m$,
  with $\Lambda = 80$ TeV, $\tan\beta = 30$  and ${\mathrm{sign}}(\mu)> 0$.}
\label{fig:decaylength}
\end{figure}

Is such models the gravitino is the lightest supersymmetric particle
and the next to LSP is commonly the lightest neutralino, $\tilde{\chi}^0_1$,
and it usually decays into a gravitino via the following channels
\begin{eqnarray}
\tilde{\chi}^0_1 &\rightarrow& \tilde{G} + \gamma\,; \nonumber\\
\tilde{\chi}^0_1 &\rightarrow& \tilde{G} + Z^0\,; \\
\tilde{\chi}^0_1 &\rightarrow& \tilde{G} + h^0\,. \nonumber
\end{eqnarray}
Due to the smallness of the neutrino-gravitino coupling, the lightest
neutralino may have a lifetime long
enough to decay far from the primary vertex of LHC generating a displaced vertex. 
We show in figure~\ref{fig:decaylength} the neutralino decay length distribution for
several values of $M_m$, for $\Lambda = 80$ TeV, $(\mu) > 0$ and
$\tan\beta = 30$. Therefore, we can anticipate
that the neutralino decay vertex can be observed at the LHCb within a large
fraction of the parameter space. \smallskip

\section{Analysis framework and results}
\label{ana:frame}

Our analysis aim to study the LHCb experiment potential to probe the
lightest neutralino decays exploring its detached vertex signature. 
We simulated the
SUSY particle production using PYTHIA version
6.4.23~\cite{Sjostrand:2000wi,Sjostrand:1993yb} where all the
properties of GMSB models were included using the SLHA
format~\cite{Skands:2003cj}. The relevant masses, mixings, branching
ratios and decay lengths were generated using the SPHENO
code~\cite{Porod:2003um,Porod:2011nf}. \smallskip

In our studies we followed the same basic strategy presented
in~\cite{decampos:2007bn}. We used a toy calorimeter roughly inspired by the
actual LHCb detector. 
We assumed a front-end detector with a pseudo-rapidity coverage
between $\eta = 1.8$ and $\eta = 4.9$. The calorimeter resolution was included
by smearing the energies with an error
\[
   \frac{\Delta E}{E} = \frac{0.10}{\sqrt{E}} \oplus 0.01 \;,
\]
for leptons. For jets with pseudo-rapidity $\eta > 3$ we have used
\[
   \frac{\Delta E}{E} = \frac{1.0}{\sqrt{E}} \oplus 0.10 \;,
\]
while for $\eta < 3$ we considered
\[
\frac{\Delta E}{E} = \frac{0.5}{\sqrt{E}} \oplus 0.03 \;,
\]
The calorimeter segmentation was assumed to be $\Delta \eta \otimes \Delta
\varphi = 0.10 \times 0.098$.
Jets were reconstructed using the cone algorithm in the subroutine
PYCELL with $\Delta R = 0.7$ and jet seed with a minimum transverse
energy $E_{T,min}^{cell} = 2$ GeV. \smallskip

Our analyzes start by selecting events that pass some typical triggers
employed by the ATLAS/CMS collaborations, {\em i.e.}  an event to be
accepted should fulfill  at least one of the following requirements:
\begin{itemize}

\item the event contains one electron or photon with $p_T > 20$ GeV;

\item the event has an isolated muon with $p_T > 6$   GeV;

\item the event exhibits two isolated electrons or photons with $p_T > 15$ GeV;

\item the event has one jet with transverse momentum in excess of 100 GeV;

\item the events possesses missing transverse energy greater than 100 GeV.

\end{itemize}

We then require the existence of, at least, one displaced vertex that
is more than $5\sigma$ away from the primary
vertex -- that is, the detached vertex is
outside the ellipsoid
\begin{equation}
  \label{eq:minellip}
      \left ( \frac{x}{5\delta_{xy}} \right )^2
   +  \left ( \frac{y}{5\delta_{xy}} \right )^2
   +  \left ( \frac{z}{5\delta_{z}} \right )^2   = 1 \; ,
\end{equation}
where the $z$-axis is along the beam direction. We used the ATLAS
expected resolutions in the transverse plane ($\delta_{xy} = 20~\mu$m)
and in the beam direction ($\delta_z = 500~\mu$m). To ensure a good
reconstruction of the displaced vertex we further required that the
lightest neutralino decays within the tracking system {\em i.e.}  
within a radius of
$500$ mm and $z$--axis length of $500$ mm. In our model the decay
lengths are such that this last requirement is almost automatically
satisfied; see figure~\ref{fig:decaylength}. \smallskip

We studied the displaced vertices coming from the neutralino decay
into $\tilde{G} + Z^0$ with the $Z^0$ subsequently decaying into either a
lepton pair or a quark pair. Usually, the $\tilde{G} + \gamma$ decay mode has
an on-shell photon and does not produce any displaced vertices, while
the $~G + h^0$ mode has a very small branching ratio in comparison
with our main $Z^0$ mode. Therefore, our main signal contains a
displaced vertex either with a oposite sign lepton pair our a jet pair
with high invariant mass. In addition to the basic cuts described above we
further required the lepton and jet pair have an invariant mass
\begin{equation}
  M_{inv} > 20 \hbox{ GeV }\; ,
\end{equation}
to avoid possible Standard Model (SM) backgrounds. In the case of a lepton pair 
signal, we further require the both lepton are isolated within a cone of $\Delta R = 0.4$.
In this sense, we
believe that, apart from instrumental provoked displaced vertices, our
signal is essencially background free.

\begin{figure}
\includegraphics[width=0.48\textwidth]{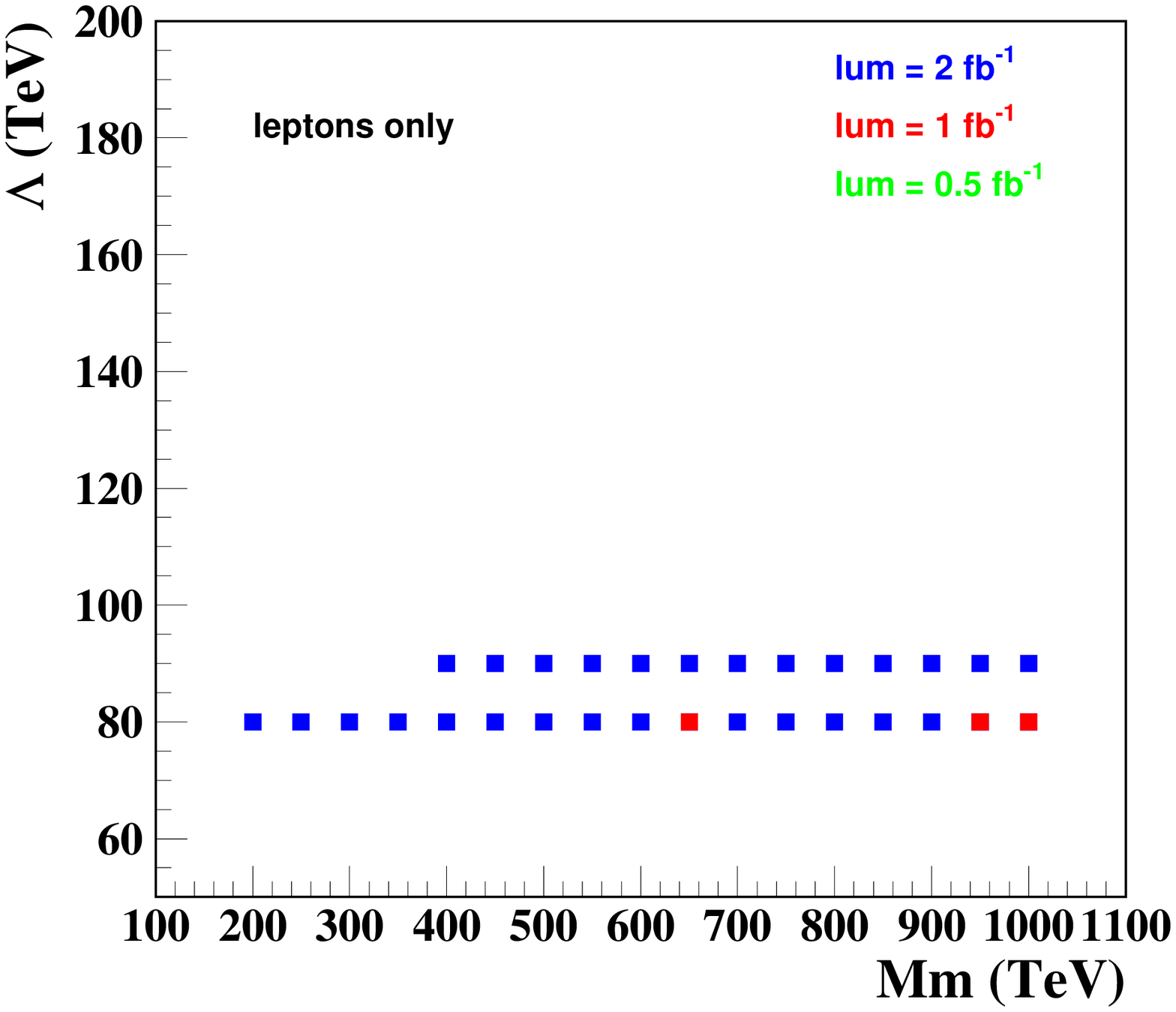}
\hfill
\includegraphics[width=0.48\textwidth]{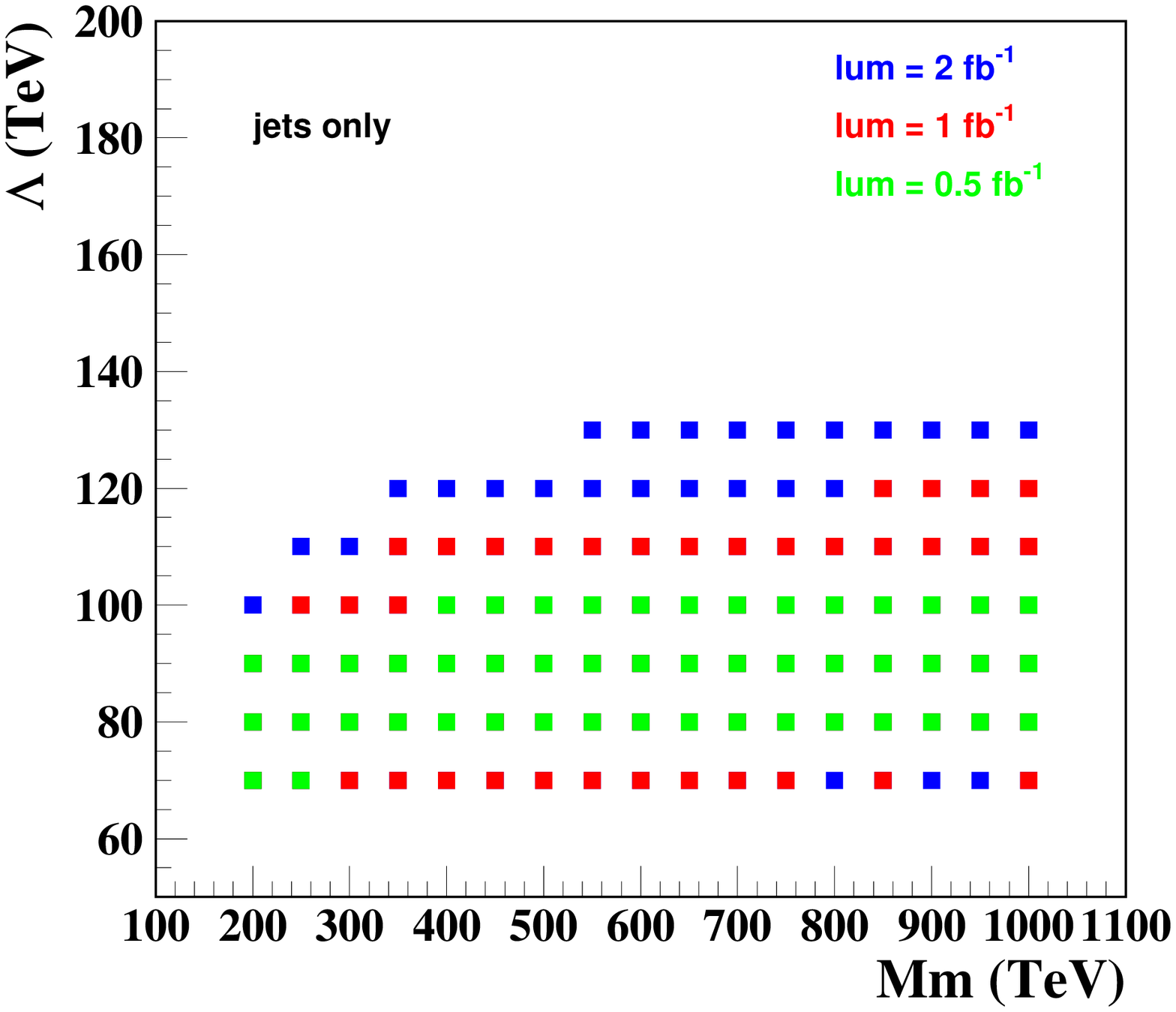}
\caption{ LHCb reach for detecting displaced vertices in the plane
  $\Lambda\times M_m$. The left panel presents the reach for detecting 
  displaced vertices including only leptons in the final state, 
  while the right one stands for the reach with only jets in the final 
  state. The blue(red)(green) squares stand for
  luminosities of $2(1)(0.5) \mathrm{fb}^{-1}$. In this figure we
  considered $\tan \beta = 30$, and $\hbox{sgn}(\mu) > 0$.}
\label{fig:reach_jl}
\end{figure}

\begin{figure}
\includegraphics[width=0.49\textwidth]{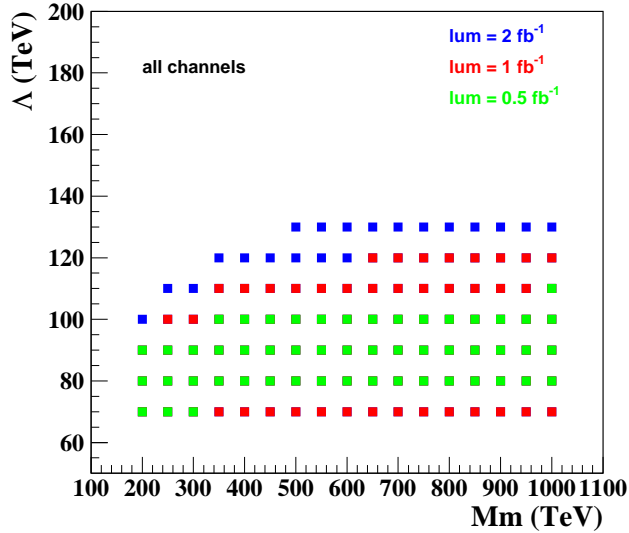}
\caption{LHCb reach for detecting displaced vertices in the plane
  $\Lambda\times M_m$.  The blue(red)(green) squares stand for
  luminosities of $2(1)(0.5) \mathrm{fb}^{-1}$. In this figure we
  considered $\tan \beta = 30$, and $\hbox{sgn}(\mu) > 0$.}
\label{gmsb_reach}
\end{figure}

In the left panel of Fig.~\ref{fig:reach_jl}, we depict the LHCb reach using only the lepton 
pair channel in the plane $\Lambda\times M_m$, with $\tan\beta = 30$ and $\mathrm{sign}(\mu) >
0$ for different luminosities. The smallness of this signal is due to the lepton isolation 
requirement which 
cuts out most of the signal. On the other hand, the jet recostruction is less restrictive and
the LHCb sensitiveness to this channel is higher, as we can observe from the right panel 
of Fig.~\ref{fig:reach_jl}.
Fig.~\ref{gmsb_reach} shows the combined LHCb reach, using both lepton and jet pair signals,
in the plane
$\Lambda\times M_m$, with $\tan\beta = 30$ and $\mathrm{sign}(\mu) >
0$ for different luminosities. We can see that for luminosities $> 1
\mathrm{fb}^{-1}$ the LHCb will be capable of detect displaced
vertex signal compatible with GMSB models with breaking scales up to 
$\Lambda \sim 130$ TeV.

\section{Conclusions}
\label{sec:conclusions}

We have analyzed the LHCb potential to detect the lightest neutralino
decay via displaced vertices in a scenario with GMSB models. 
In such models, for a large range of values in the plane
$\Lambda\times M_{m}$, one can have sizeable
displaced vertices containing the final states $\ell^+\ell^-,\,
(\ell=e,\mu,\tau)$ and di-jets for luminosities that the LHCb will
reach in a near future. We saw that the LHCb is capable of probe such 
models up to $\Lambda < 130$ TeV with a small dependence of the messenger 
mass $M_m$.

Recent Higgs search results has been performed by ATLAS and CMS 
experiments at CERN \cite{higgs:search} with a result of a Higgs boson with a 
mass of $125$ GeV. However, further studies show that the 
properties of the detected Higgs boson may not comply with a SM Higgs 
\cite{higgs:props}. In minimal GMSB models considered here, the lightest Higgs 
mass is hardly heavier than $118$ GeV. 
Nevertheless, in order to make GMSB models viable, one can think in extensions 
of the minimal GMSB model where the Higgs mass can recieve contribuitions to be 
consistent with data. For instance, one can introduce mixings among the 
messengers and matter that can enlarge the Higgs mass up to the experimental 
value \cite{higgs:albaid} even in the range of the $\Lambda\times M_m$ plane 
considered here. Although such models may change the spectroscopy of 
SUSY, specially in the third family scalar sector,  we believe their 
phenomenology would still be similiar to those considered here, since we are
mainly interested in the lightest neutralino decay.

\section*{Acknowledgments}

%
M.B.M. would like to thank the Departamento de F\'{\i}sica--Matem\'atica of 
Institute of Physics of University of S\~ao Paulo for their hospitality.



\end{document}